\documentclass[preprint2]{aastex}

\newcommand{\nustar}{{\it NuSTAR}}
\newcommand{\swift}{{\it Swift}}
\newcommand{\chandra}{{\it Chandra}}
\newcommand{\xmm}{{\it XMM-Newton}}
\newcommand{\suzaku}{{\it Suzaku}}
\newcommand{\vla}{{VLA}}

\def\arcsec{$\,^{\prime\prime}$}

\slugcomment{To be submitted in ApJ}

\shorttitle{X-ray and Radio study of V404 Cyg}
\shortauthors{Rana et al.}

\begin{document}


\title{Characterizing X-ray and Radio emission in the Black Hole X-Ray Binary V404 Cygni during Quiescence}


\author{Vikram Rana\altaffilmark{1}, Alan Loh\altaffilmark{2}, Stephane Corbel\altaffilmark{2,3}, John A. Tomsick\altaffilmark{4}, Deepto Chakrabarty\altaffilmark{5}, Dominic J. Walton\altaffilmark{1}, Didier Barret\altaffilmark{6}, Steven E. Boggs\altaffilmark{4}, Finn E. Christensen\altaffilmark{7}, William Craig\altaffilmark{4,8}, Felix
Fuerst\altaffilmark{1}, Poshak Gandhi\altaffilmark{9}, Brian W. Grefenstette\altaffilmark{1}, Charles Hailey\altaffilmark{10}, Fiona A. Harrison\altaffilmark{1}, Kristin K. Madsen\altaffilmark{1}, Farid Rahoui\altaffilmark{11,12}, Daniel Stern\altaffilmark{13}, Shriharsh Tendulkar\altaffilmark{1}, William W. Zhang\altaffilmark{14}}

\altaffiltext{1}{Cahill Center for Astronomy and Astrophysics, California Institute of Technology, Pasadena, CA 91125}
\altaffiltext{2}{Laboratoire AIM (CEA/IRFU - CNRS/INSU - Universit\'e Paris Diderot), CEA DSM/IRFU/SAp, F-91191 Gif-sur-Yvette, France}
\altaffiltext{3}{Station de Radioastronomie de Nan\c{c}ay, Observatoire de Paris, PSL Research University, CNRS, Univ. Orl\'eans, OSUC, 18330 Nan\c{c}ay, France}
\altaffiltext{4}{Space Sciences Laboratory, University of California, Berkeley, CA 94720-7450, USA}
\altaffiltext{5}{MIT Kavli Institute for Astrophysics and Space Research, MIT, Cambridge, MA 02139, USA}
\altaffiltext{6}{Universit\'e de Toulouse; UPS-OMP; IRAP; Toulouse, France}
\altaffiltext{7}{DTU Space, National Space Institute, Technical University of Denmark, DK-2800 Lyngby, Denmark}
\altaffiltext{8}{Lawrence Livermore National Laboratory, Livermore, CA 94550, USA}
\altaffiltext{9}{Department of Physics, Durham University, Durham, DH1 3LE, UK}
\altaffiltext{10}{Columbia Astrophysics Laboratory, Columbia University, New York, NY 10027, USA}
\altaffiltext{11}{European Southern Observatory, Karl Schwarzschild-Strasse 2, 85748 Garching bei Munchen, Germany}
\altaffiltext{12}{Department of Astronomy, Harvard University, 60 Garden Street, Cambridge, MA 02138, USA}
\altaffiltext{13}{Jet Propulsion Laboratory, California Institute of Technol- ogy, Pasadena, CA 91109, USA}
\altaffiltext{14}{NASA Goddard Space Flight Center, Greenbelt, MD 20771, USA}



\begin{abstract}
We present results from multi-wavelength simultaneous X-ray and radio observations of the black hole X-ray binary V404~Cyg in quiescence. Our coverage with \nustar\ provides the very first opportunity to study the X-ray spectrum of V404~Cyg at energies above 10 keV. The unabsorbed broad-band (0.3--30 keV) quiescent luminosity of the source is 8.9$\times$10$^{32}$ erg s$^{-1}$ for a distance of 2.4 kpc. The source shows clear variability on short time scales (an hour to a couple of hours) in radio, soft X-ray and hard X-ray bands in the form of multiple flares.  
The broad-band X-ray spectra obtained from \xmm\ and \nustar\ can be characterized with a power-law model having photon index $\Gamma$=2.12$\pm$0.07 (90\% confidence errors); however, residuals at high energies indicate spectral curvature significant at a 3$\sigma$ confidence level with e-folding energy of the cutoff to be 20$^{+20}_{-7}$ keV. 
Such curvature can be explained using synchrotron emission from the base of a jet outflow.
Radio observations using the VLA reveal that the spectral index evolves on very fast time-scales (as short as 10 min.), switching between optically thick and thin synchrotron emission, possibly due to instabilities in the compact jet or stochastic instabilities in accretion rate. We explore different scenarios to explain this very fast variability. 

\end{abstract}


\section{Introduction}
Low-mass X-ray binaries (LMXBs) are systems containing a compact object, either a black hole (BH) or a neutron star (NS), accreting material from a low mass (less than a solar mass) late-type companion star. These systems often show transient behavior with long intervals (years to decades) of low luminosity emission (namely the quiescent state) between bright outburst states, lasting from weeks to months. X-ray luminosities can typically have values in the range of 10$^{31-33}$ erg s$^{-1}$ during quiescence, dramatically increasing by several orders of magnitude during outbursts. A disk instability model, originally proposed to explain dwarf nova outbursts, can broadly explain outbursts in LMXBs \citetext{\citealp{cannizzo93, lasota00}}. Several models have been proposed to explain the origin of the quiescent X-ray emission in these systems. During the low-luminosity phase the accretion rate from the companion drops below a critical value, resulting in drastic structural changes in the accretion disk. The inner part of the disk can evaporate forming an advection-dominated accretion flow (ADAF; \citealp{narayan94}) that is radiatively inefficient. Another possibility invokes a jet-dominated phase where a significant amount of energy is released in an outflow instead of in an ADAF \citep{fender03}. \citet{yuan05} suggested another interesting possibility of a hybrid of jet and ADAF models to explain spectral and timing characteristics of XTE J1118+480. \citet{blandford99} proposed a modified version of ADAF in which powerful winds carry away mass, angular momentum and energy from the accreting gas. Typically, it has been presumed that accretion onto a quiescent BH happens via some kind of radiatively inefficient accretion flow where X-rays are emitted via a population of hot electrons. However, it is still not clear if the quiescent BH always produces steady collimated jets. Owing to the quiescent nature of these BH systems, it is difficult to obtain good quality multi-wavelength data, with V404~Cyg being one of the few sources with a high-quality broad-band SED in the quiescent state.

V404 Cyg has been used as a primary test bed for many of the above models owing to its close proximity and high luminosity during quiescence \citetext{\citealp{narayan96, narayan97}}. X-ray emission was discovered from V404 Cyg during its outburst in 1989 by the $Ginga$ X-ray satellite \citep{makino89}. Before that, it had undergone two outbursts, in 1938 and 1956, which were only seen in optical due to lack of space instruments. 
Recently, the source has been observed in outburst for
a fourth time when it was detected by {\em Swift}/BAT on 
2015 June 15 \citep{barthelmy15}. Following the $Swift$ BAT trigger, the source has been extensively monitored at multi-wavelengths including optical, radio, UV and X-rays \citetext{see e.g. \citealp{gazeas15}, \citealp{gandhi15}, \citealp{kuulkers15} and several other related ATels.}. V404~Cyg is believed to contain a K0($\pm$1) III-V secondary star orbiting around a $\sim$10 M$_{\odot}$ BH \citep{shahbaz96} with an orbital period of 6.47 days. More recently, \citet{khargharia10} estimated a slightly lower value of 9$^{+0.2}_{-0.6}$ M$_{\odot}$ for the BH using near-infrared spectroscopy. At a distance of 2.39$\pm$0.14 kpc, obtained using parallax measurement \citep{miller09}, V404~Cyg contains the most luminous quiescent BH, brighter by about two orders of magnitude \citep{garcia01}. Typically, V404 Cyg is known to vary in X-rays by a factor of 2--10 on time scales of a few hours in quiescence \citetext{\citealp{wagner94}; \citealp{kong02}; \citealp{bernardini14}}. However, \citet{hynes04} reported X-ray flux variation with unusually large amplitude (factor of $ \gtrsim $20 in 60 ks). The X-ray variations are well correlated with that observed in the optical H$_{\alpha}$ line, suggesting that the optical variations are powered by X-ray irradiation. Strong variability has also been reported at radio wavelengths, represented as large flares and dips \citep{hynes09} with no clear correlations between X-ray and radio emission levels. 

In this paper we present results from our recent X-ray and radio observations of V404~Cyg in the quiescent state. The next section describes the observations and data analysis for \xmm, \nustar\ and the \vla. Sections~3 and 4 present results from timing and spectral analyses.
We discuss results in Section~5 and summarize them in Section~6.

\section{Observations and Data Reduction}
We observed V404~Cyg with \xmm, \nustar\ ($Nuclear$ $Spectroscopic$ $Telescope$ $Array$) and the \vla\ (Very Large Array) as part of a multi-wavelength campaign to study its broadband properties in quiescence. 
\nustar\ \citep{harrison13} has two coaligned hard X-ray optics modules with corresponding focal planes, referred to as FPMA and FPMB, that cover 3--79 keV energy range. \xmm\ employs X-ray optics with PN and MOS CCD detectors that cover the 0.3--10 keV energy range.
The \vla\ provides coverage in the radio part of the electromagnetic spectrum in 4.74--7.96 GHz.  
V404~Cyg was observed simultaneously with \nustar\ and \xmm\ on 2013 October 13-14 for about 150 ks and 52 ks, respectively. According to our original observation strategy, the \vla\ observations were supposed to be simultaneous with the \xmm\ and first \nustar\ observations. However, radio observations were canceled due to a \vla\ shutdown. Therefore, we obtained additional simultaneous X-ray and radio data with \nustar\ and the \vla\ on 2013 December 2
with about 25 ks exposure for \nustar. Details of the X-ray observations are listed in Table~\ref{obs-log}.

\begin{deluxetable}{ccccc}
\tabletypesize{\scriptsize}
\tablecaption{X-ray observation log for V404~Cyg. \label{obs-log}}
\tablewidth{0pt}
\tablehead{
\colhead{Observatory} & \colhead{ObsID} & \colhead{Start time} & \colhead{Stop time} & \colhead{Exposure} \\
\colhead{ } & \colhead{ } & \colhead{UTC} & \colhead{UTC} & \colhead{(ks)}
}
\startdata
\nustar\ (Epoch 1) & 30001010002 & 2013-­10-­13 12:06:07 & 2013-10-14 21:46:07 & 61.4    \\
\nustar\ (Epoch 2) & 30001010003 & 2013-­10-­14 21:46:07 & 2013-10-16 23:16:07 & 90.9     \\
\nustar\ (Epoch 3) & 30001010005 & 2013-­12-­02 17:36:07 & 2013-12-03 06:21:07 & 24.9    \\
\xmm\    & 0723310201  & 2013-­10-­13 18:23:24 & 2013-10-14 07:38:02  & 52.6   \\
\enddata
\end{deluxetable}

\begin{figure}
\includegraphics[angle=-90,scale=.32]{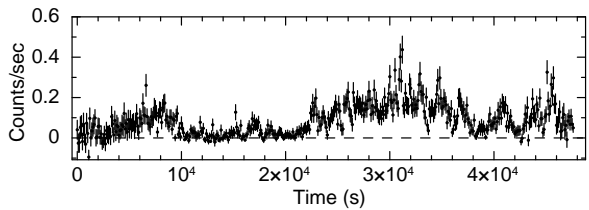}
\caption{\xmm\ pn light curve of V404 Cyg in the 0.5--10 keV energy band {\bf with 100 s time binning}.
\label{xmm-lc}}
\end{figure}

\begin{figure}
\includegraphics[angle=-90,scale=.32]{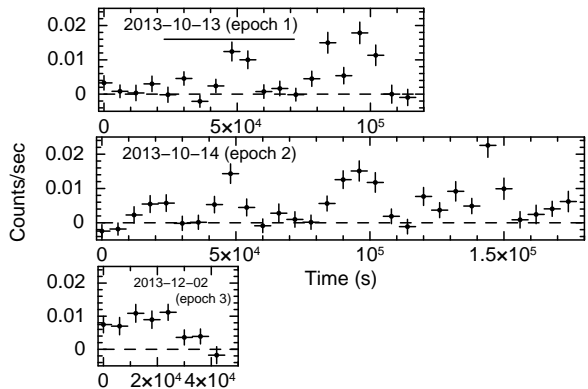}
\caption{\nustar\ FPMA light curves for three epochs of observations in the 3--25 keV energy band with 6000 s time binning.
The solid horizontal line in the top panel indicates the time of the \xmm\ observation. 
\label{nu-lc}}
\end{figure}

\begin{figure}
	\includegraphics[angle=0,scale=0.43]{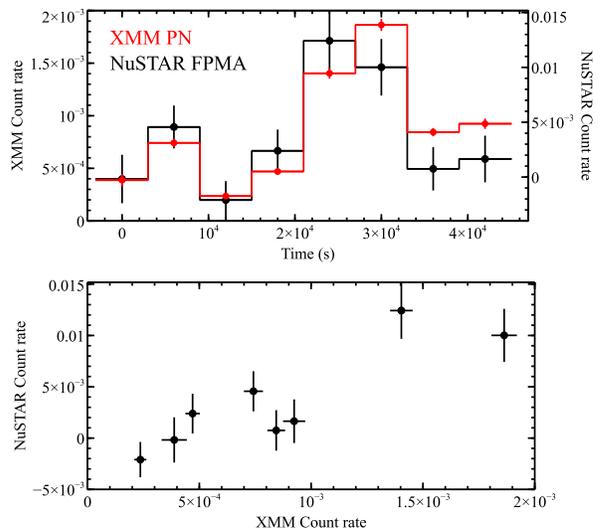}
	\caption{\xmm\ and \nustar\ light curves (top panel) for the overlapping section of the observation (horizontal line in Figure~{\ref{nu-lc}} top panel)  with 6000 s time binning. The two count rates are plotted against each other (bottom panel) to see correlation between the two light curves.}
	\label{xmm-nu-lc} 
\end{figure}

\subsection{XMM-Newton}
We obtained a
52.6\,ks observation of V404 Cyg with {\em XMM-Newton} (ObsID 0723310201) 
on 2013 October 13-14, covering part of the first {\em NuSTAR} observation (see Table~\ref{obs-log}).  
We used the Scientific Analysis System (SAS) v13.0.1 software package 
to reduce the data from the pn, MOS1, and MOS2 instruments, which were all
operated in full frame mode with the medium blocking filter.  The 10--12\, keV
light curve from the full pn detector shows that proton flares occurred
at the beginning and end of the observation, and we filtered out the 
time when 10--12\, keV pn count rate was greater than 1\,c/s, leaving
a net exposure time of 31.5\,ks for pn and 34.9\,ks for each of the MOS
units.  The part of the observation used to make energy spectra started
at 2013 October 13, 20.2 h UT and ended at 2013 October 14, 6.0 h UT.

We produced event files using {\ttfamily epproc} and {\ttfamily emproc}, 
and then applied the standard event filtering.  For pn, this means keeping
events with the ``FLAG'' parameter set to zero, the ``PATTERN'' parameter 
less than or equal to 4, and applying the \#XMMEA\_EP data selection.  For
the MOS units, the filtering criterion is a ``PATTERN'' parameter less 
than or equal to 12 along with applying the \#XMMEA\_EM data selection.
The pn and MOS images clearly show the presence of a point source at the
known location of V404~Cyg, which is the brightest source in the field.

We extracted a 0.5--10\,keV pn light curve, a 0.5--10\,keV pn spectrum and 
0.3--10\,keV MOS spectra using a circular extraction region centered on 
V404~Cyg with a radius of 720 pixels (0.6 arcminutes).  We used nearby 
source-free rectangular regions to estimate and subtract the contribution
from background.  We rebinned the pn spectrum to have at least 100 counts
per bin and the MOS spectra to have at least 50 counts per bin.  We produced
pn and MOS response matrices with {\ttfamily rmfgen} and {\ttfamily arfgen}.
The average count rates for pn, MOS1, and MOS2 are $0.105\pm 0.002$\,c/s, 
$0.036\pm 0.001$\,c/s, and $0.035\pm0.001$\,c/s, respectively. 

\subsection{NuSTAR}
We reduced the \nustar\ data using the \nustar\ 
Data Analysis Software (NuSTARDAS v1.4.1) available as part of HEASOFT v6.16 and the latest CALDB files. 
The \nustar\ observations were taken during intervals of normal Solar activity,
and we used standard filtering to remove periods of high background during South Atlantic Anomaly (SAA) passages and
Earth occultation.   We created cleaned, calibrated event files using the {\tt NUPIPELINE} script with standard settings.
The source spectra and light curves were extracted using a 50\arcsec\ radius circular region around the source. For background estimation on FPMA, we used {\it nuskybgd} tool that accurately generates the background by properly taking into account the non-uniformity of background across the focal plane \citetext{see \citealp{wik14} for details}.
 
FPMB, on the other hand, showed the presence of faint stray light leaking into the field-of-view from a nearby source during epochs 1 and 2. The {\it nuskybgd} tool does not account for stray light; thus we did not use it for FPMB. Fortunately, for epoch 1, the source fell on a favorable position and we were able to remove stray light contamination by carefully selecting the background from a 100\arcsec\ radius circular region and subtracting it from the source. However, for epoch 2 the source was right on the edge of the stray light pattern, making it difficult to remove the contamination. Therefore, to safely avoid any contamination issue, we did not use epoch 2 FPMB data for any further analysis. FPMB epoch 3 data and all FPMA data are clean and free from any stray light issue and we use them for detailed scientific analysis.

\subsection{VLA}
V404~Cyg was observed with the upgraded Jansky \vla\ \citep[project 13B--016, PI: S.~Corbel]{2011ApJ...739L...1P} on 2013 December 2 for a total time on source of $\sim$  $9\,$hours.
The array was in the B-configuration and the observations were performed in the C band ($4.74 - 7.96\,$GHz) with two $1024\,$MHz base-bands centered at $5.25$ and $7.45\,$GHz, with eight spectral windows of $64 \times 2\,$MHz channels for each baseband. 

The quasar 3C286 (J1331$+$3030) was used for the bandpass calibration and to set the amplitude scale.
We used the very nearby compact source J2025$+$3343 ($0\fdg28$ away from V404~Cyg) to calibrate the antenna amplitude and phase gains.
We performed a cycle of $10$ minutes on V404~Cyg and $36$ seconds on the phase-referenced secondary calibrator.
Two additional 3-minutes scans were allocated on the non-polarized compact source J2355$+$4950 to calibrate the polarization leakages.
Flagging, calibration and imaging of the data followed standard procedures and were carried out within the Common Astronomy Software Application (CASA, version 4.1.0) package \citep{2007ASPC..376..127M}.

\begin{figure}
\includegraphics[angle=0,scale=.40]{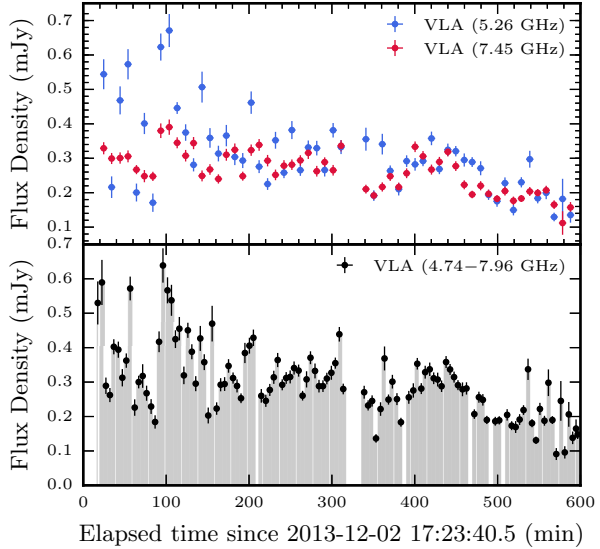}
\caption{V404 Cyg light curves computed over \vla\ radio frequency bands as labeled in each panel. The binning times correspond to 1 and 0.5 scan duration ($\sim$10 min) for the upper and bottom panels respectively.
\label{vla-lc}}
\end{figure} 
 
We set the primary calibrator 3C286 flux scales using the task \texttt{setjy} and the physical properties measured by \citet{2013ApJS..206...16P}.
The routines \texttt{bandpass} and \texttt{gaincal} were used to solve for the complex bandpass, the amplitude, and the phase gains.
Stokes parameter values of the linearly-polarised calibrator 3C286 were assigned with \texttt{setjy} thanks to its fractional linear polarisation and its polarisation position angle listed in \citet{2013ApJS..206...16P}.  
We derived calibration solutions for the leakage terms using J2355$+$4950.
The flux densities of the secondary calibrators were bootstrapped using the task \texttt{fluxscale}, based on the absolute calibration of 3C286.
Finally, we applied the calibration to V404~Cyg, linearly interpolating the gain solutions from the secondary calibrator J2025$+$3343, assuming that the two sources are close enough so that the gains are slowly varying between the two nearby field of views.
 
We imaged V404~Cyg in all Stokes parameters using the Cotton-Schwab CLEAN algorithm for the deconvolution.
Calibrated measurements were split into four distinct $512\,$MHz bands in order to characterize the spectral evolution of  V404~Cyg over the bandpass. 
We made $480 \times 480\,$pixel$^2$ images with $0\farcs15$ pixel width. At $7.94\,$GHz the maximum baseline is $\sim$270k$\lambda$, corresponding to an angular scale of 
$\sim$ $0\farcs76$. 
We used the multi-frequency synthesis method with the Briggs' weighting robust parameter set to $0.5$ for all $I$ images in order to achieve the best compromise between  angular resolution and  signal-to-noise ratio.
However, the $U$ and $Q$ images were deconvolved with natural weighting to maximize the S/N ratio.  
V404~Cyg is consistent with a point source in all images, so the flux densities were measured by fitting a point source in the image plane, and we add a $1$\% systematic error as usual for the \vla\ in this frequency range.

\begin{figure}
\includegraphics[angle=-90,scale=.35]{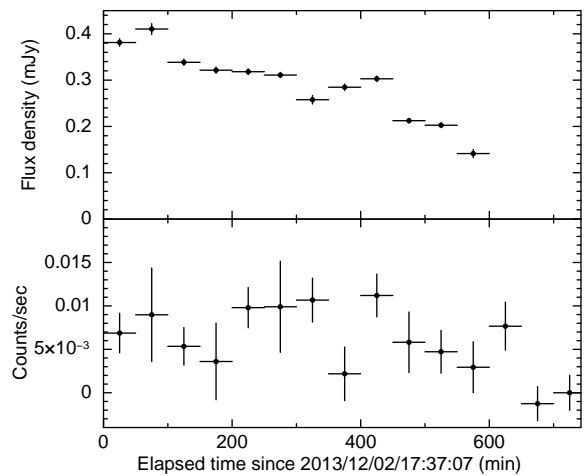}
\caption{Radio (top panel) and X-ray (bottom panel) light curves from simultaneous observations with \vla\ and \nustar\ during epoch 3 with 3000 s time bin. The X-ray light curve is in 3--10 keV energy band and radio in 4.74--7.96 GHz. 
\label{radiox-lc}}
\end{figure}

\begin{figure}
	\includegraphics[angle=0,scale=0.5]{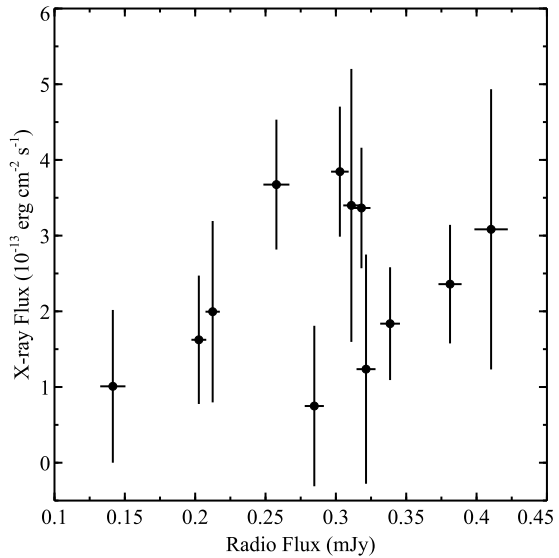}
	\caption{3--10 keV X-ray flux from \nustar\ epoch 3 observation and radio flux (4.74--7.96 GHz).
		\label{radiox_corr}}
\end{figure}

\subsection{Radio Flux and Polarization Measurement}
V404~Cyg is detected with a mean flux density of $300 \pm 2\,\mu$Jy (Stokes $I$ over the full bandwidth) in good agreement with previous studies conducted during the quiescent state \citep[e.g.,][]{hynes09, 2008MNRAS.388.1751M, 2005MNRAS.356.1017G}.
The source is unresolved with a beamsize of $0.81 \times 0.79\,$arcsec$^2$ at $7.7\,$GHz, consistent with previous high spatial resolution studies \citep{2008MNRAS.388.1751M}.

No significant polarised emission is detected at the location of V404~Cyg, as no point source can be fitted on the images in the Stokes $U$, $Q$ and $V$ parameters.  
By extracting the flux from a reduced number of channels, we avoided smearing out of any polarised signal over a large frequency bandwidth due to Faraday rotation of the linear polarisation vector. 
Therefore, we computed the upper limit on its polarised emission by taking into account the rms of the corresponding images, i.e. $1.51$, $1.48\,$ and $1.47\,\mu$Jy for Stokes parameters $Q$, $U$ and $V$, respectively.
The linear polarisation, defined as $LP = \sqrt{Q^2 + U^2}$, can then be constrained not to exceed $LP \leq 6.33\,\mu$Jy at a $3\sigma$ confidence level.
Regarding the fractional polarisation $FP = 100 LP / I$, an upper limit of 2.11\% can be inferred. Similar amplitude measurements have been obtained for black hole candidates in brighter, hard states (e.g. GX~339-4; \citealp{2000A&A...359..251C}), however, it is close to the detection limit. 

\begin{figure}
\includegraphics[angle=-90,scale=.32]{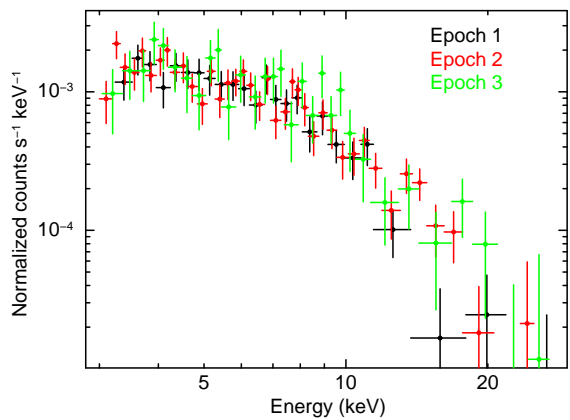}
\caption{\nustar\ FPMA spectra for three epochs of observations. Black, red and green points correspond to epochs 1, 2, and 3, respectively.
\label{nu-spec}}
\end{figure}

\section{Timing Analysis}
Figure~\ref{xmm-lc} shows the background subtracted X-ray light curve of V404~Cyg in the 0.5--10 keV energy band from \xmm\ pn with 100~s time binning. The light curve clearly shows significant variability due to flaring activity from the source. The count rate shows variability by a factor of 4--5 during the \xmm\ observation on a time scale of a few hours. Figure~\ref{nu-lc} shows hard X-ray light curves in the 3--25 keV energy band from \nustar\ FPMA module. The three panels correspond to the three epochs of \nustar\ observations, with a solid horizontal line in the top panel indicating the interval corresponding to the \xmm\ observation. \nustar\ light curves are binned to
6000~s, approximately corresponding to an orbit of \nustar. The source clearly shows significant variability in the hard X-ray band, with similar time scale as \xmm. This suggests that the flaring behavior continues in the hard X-ray band as well. During the epoch 3 snap-shot observation with \nustar, the source was caught mainly in a flaring state, suggesting that it was showing persistent flaring behavior during the entire span of these observations. The three epochs of \nustar\ observations caught about 6--7 small flares from this source. Figure~\ref{xmm-nu-lc} shows X-ray light curves from the overlapping sections of the \xmm\ and \nustar\ epoch 1 observations (top panel) to facilitate a direct comparison between the soft X-ray (0.5--10 keV) and hard X-ray (3-25 keV) variations. The two light curves show very similar profile suggesting a strong correlation between the two X-ray bands. We plot the \xmm\ count rate and the \nustar\ count rate in the bottom panel of Figure~{\ref{xmm-nu-lc}} to look for any correlation between the two light curves in top panel. It is clearly visible that the X-rays in two bands are correlated. In order to quantify this correlation, we measured the Pearson's correlation coefficient between the two light curves to be 0.9$^{+0.1}_{-0.2}$ (90\% confidence errors). All the errors in this paper are quoted with 90\% confidence limit for a single parameter of interest, unless otherwise noted.  A value of 1 for this coefficient indicates a strong positive correlation and a value of 0 indicates no correlation. Hence, a value of 0.9 suggests a strong correlation between 0.5--10 keV and 3--25 keV X-rays, as expected.

In order to search for shorter time scale variability in radio, the calibrated \vla\ data were split into time segments with a duration chosen to be as small as possible for the detection to be considered significant. 
The source was fit in the image plane and the resulting radio light curves of V404~Cyg are shown in Figure~\ref{vla-lc}, 
highlighting the flux density variations across the observation.
V404~Cyg shows evidence of fast radio variability with a flux density ranging from $0.09$ to $0.64\,$mJy (integrated over the whole frequency band). There is the suggestion of a small flaring activity around $100\,$min after the beginning of the observation, with a $0.64\,$mJy peak level. 
This flare is characterized by a fast rise of less than $5\,$min, followed by a slower decrease over $\sim 30\,$min.

In order to directly compare the source variability in the radio and X-ray bands, we plot their respective light curves in Figure~\ref{radiox-lc}. The light curves are from the  simultaneous radio (\vla\ in 4.74--7.96 GHz) and X-ray observations (\nustar\ epoch 3 in the 3--10 keV energy band) and binned to 3000 s. The source exhibits significant variability at both radio and X-ray energies. The radio flux density changes by a factor of $\sim$3 during the time of observation. Although clear variability is present in both bands, there is no strong correlation between radio and X-ray fluxes, partly due to the poor signal-to-noise ratio of the X-ray data. Once again we measured the Pearson's correlation coefficient between radio and X-ray time series, to quantify any correlation between the two. The correlation coefficient between radio and X-ray bands is  
0.4$^{+0.2}_{-0.3}$. A value of 0.4 suggests a weak positive correlation, though the large errors make it statistically insignificant. Figure~\ref{radiox_corr} shows X-ray and radio flux variation for better visualization. Thus we conclude that there is no compelling evidence for correlation between the simultaneous radio and X-ray data analyzed here. 

\begin{figure}
\includegraphics[angle=-90,scale=.38]{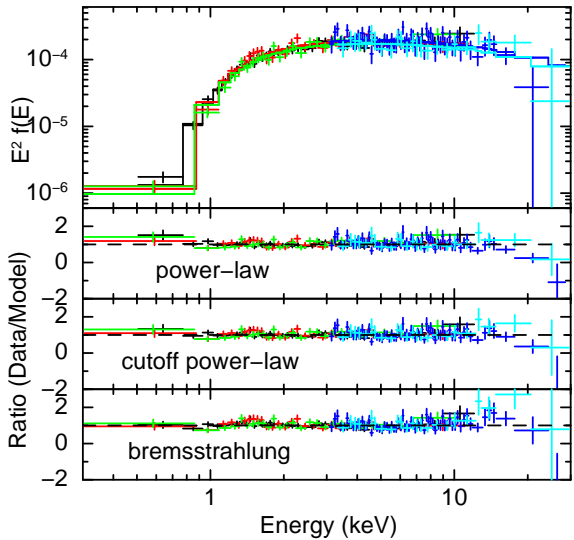}
\caption{Top panel: Broad-band V404~Cyg spectra from \xmm\ (EPIC-pn in black, MOS1 in red and MOS2 in green) and \nustar\ (FPMA in blue and FPMB in light blue). The solid lines represent best fit power-law models. The lower panels show the ratio of the data to different spectral models, as labeled.   
\label{3model_ratio}}
\end{figure}

\begin{deluxetable}{ccccccc}
\tabletypesize{\scriptsize}
\tablecaption{Best fit spectral parameters from joint fit to \xmm\ and \nustar\ data. \label{spec-param}}
\tablewidth{0pt}
\tablehead{
\colhead{Model} & \colhead{$N_{\rm H}$} & \colhead{$\Gamma$/$\tau_p$} & \colhead{$E_{cut}$/kT$_e$} & \colhead{Norm} & \colhead{$\chi^2$/dof} & \colhead{Flux\tablenotemark{*}} \\
\colhead{ } & \colhead{($10^{22}$ cm$^{-2}$) } & \colhead{ } & \colhead{(keV)} & \colhead{(10$^{-4}$) } &
\colhead{} & \colhead{}
}
\startdata
Power-law & 1.25$\pm$0.08 & 2.12$\pm$0.07 & --- & 2.20$\pm$0.20 & 173/200 & 3.31  \\
Cut-off power-law & 1.14$\pm$0.10 & 1.86$\pm$0.15 & 20$^{+20}_{-7}$ & 1.92$^{+0.22}_{-0.20}$ & 161/199 & 3.39   \\
Bremsstrahlung & 0.84$\pm$0.05 & --- & 6.9$^{+0.7}_{-0.6}$ & 1.58$^{+0.08}_{-0.07}$ & 199/200 & 3.60  \\
CompTT & 0.78$\pm$0.06 & 4.8$^{+0.7}_{-0.8}$ & 4.0$^{+1.2}_{-0.6}$& 0.60$^{+0.11}_{-0.13}$ & 160/199 & 3.42 \\
\enddata
\tablenotetext{*}{Absorbed flux in 3--10 keV energy band with units of 10$^{-13}$ erg cm$^{-2}$ s$^{-1}$.}
\end{deluxetable}

\section{Spectral Analysis}
\subsection{X-ray Spectral Analysis}
We studied the spectra for each X-ray observation separately in order to investigate any spectral variation from epoch to epoch. Throughout this work we performed spectral modeling using XSPEC v12.8.2 \citep{arnaud96}. Figure~\ref{nu-spec} shows the \nustar\ FPMA spectra for the three epochs of observations. It is readily visible that the spectra follow each other very closely, suggesting there is very little spectral variation among these three epochs. In order to quantify any spectral variation, we examined each spectra separately (three \nustar\ spectra and one \xmm ). We started with a simple model consisting of a power law modified by photoelectric absorption due to interstellar medium.   
To account for the absorption due to intervening material, we used the $tbabs$ spectral model with updated solar abundances \citep{wilms00} and photoionization cross-sections as described in \citet{verner96}. 

For \xmm, the EPIC (pn and MOS) spectra were fitted simultaneously. We obtained a best fit value of photon index of $\Gamma$=2.04$\pm$0.08 and absorption $N_{\rm H}$=(1.18$\pm$0.09) $\times$ 10$^{22}$ cm$^{-2}$. The absorbed 0.3--10 keV flux is 5.5 $\times$ 10$^{-13}$ erg s$^{-1}$ cm$^{-2}$, which agrees well with the previously reported value of 5.8 $\times$ 10$^{-13}$ erg s$^{-1}$ cm$^{-2}$ \citep{bernardini14}. The absorption obtained from \xmm\ is a little higher than the Galactic value of 8.10 $\times$ 10$^{21}$ cm$^{-2}$ \citep{dickey90}. While performing fits with individual \nustar\ spectra, we fixed $N_{\rm H}$ to the value obtained with \xmm\ as \nustar's energy range is not sensitive to such absorption values. 
The best fit values for the power-law index for the three \nustar\ observations are 2.54$^{+0.18}_{-0.17}$, 2.28$\pm$0.17 and 1.90$\pm$0.24, respectively. The spectrum apparently is flattening going from epoch 1 to epoch 3, however, the individual errors on the power-law indices overlap, suggesting that the difference may not be statistically significant. In order to investigate this we fitted the data from each epoch simultaneously with an absorbed power-law model. All the parameters were linked between  epochs and only the overall normalization constants were allowed to vary. This also provided an excellent fit to the data, with a reduced $\chi^2$ of 1.07 for 95 degrees of freedom and a best fit power-law index of $\Gamma$=2.35$\pm$0.12. This suggests that any variations in the spectral parameters during the three epochs are not significant and the spectra can formally be considered to be consistent with each other. Further, we have also derived hardness ratios between the two \nustar\ energy bands, 3--8 keV (soft) and 8--30 keV (hard) to test for any X-ray spectral variations. We found no variation in the hardness ratio (hard/soft), indicating that there are no significant spectral changes in the X-ray data. This is consistent with the previous X-ray studies using $Chandra$ \citep{hynes09} and $Swift$ \citep{bernardini14}. 

\begin{figure}
\includegraphics[scale=1.1]{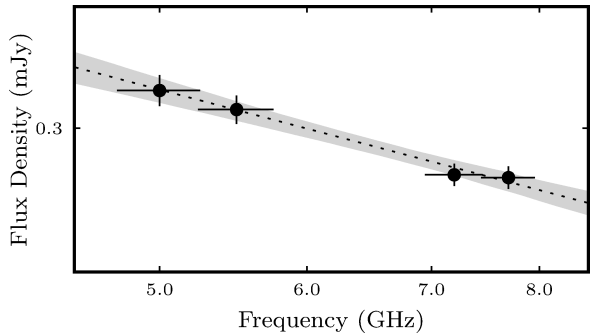}
\caption{V404~Cyg radio spectrum as described in Table~\ref{vla-param}. The dotted black line and the gray shaded area represent the fitted power-law shape spectrum and the $95\,$\% confidence interval respectively. The spectral index for the power-law is $\alpha$=-0.27$\pm$0.03.
\label{vla-spec}}
\end{figure}

In order to evaluate the broad-band X-ray characteristics of V404~Cyg during quiescence, we simultaneously fitted \xmm\ and \nustar\ data to achieve energy coverage from 0.3--30 keV. As shown above, there is no significant spectral variation between the three epochs of \nustar\ observations, therefore we combined the data from the three \nustar\ epochs to improve the signal-to-noise ratio. We used the ftool $addascaspec$ to combine the \nustar\ data, which properly combines the corresponding background and response files. In order to account for the cross-calibration between different instruments, we included a constant multiplicative factor in all the spectral modeling described below. This constant factor was held fixed to 1.0 for \xmm\ EPIC-pn and allowed to vary freely for other instruments. The EPIC-pn to \nustar\ normalization differences are below the expected $\lesssim$10\% level \citep{madsen15}.

We employed several different spectral models to characterize the broad-band spectra of V404~Cyg during quiescence, including an absorbed power-law model, a cutoff power-law model, a thermal bremsstrahlung model and a thermally comptonized continuum model (see Figure~\ref{3model_ratio}). The best fit spectral parameters obtained through joint fitting of \xmm\ and \nustar\ data are listed in Table~\ref{spec-param}. All these models provide statistically acceptable fits to the data. The power-law model provides a best fit value of the photon index of $\Gamma$=2.12$\pm$0.07. The unabsorbed flux in 0.3--30 keV energy band is (1.29$\pm$0.07) $\times$ 10$^{-12}$ erg cm$^{-2}$ s$^{-1}$ and the corresponding luminosity is (8.9$\pm$0.5) $\times$ 10$^{32}$ erg s$^{-1}$ for a distance of 2.4 kpc \citep{miller09}. Although the power-law model provides an acceptable fit ($\chi^2_{\nu}$=0.86), the residuals show a hint of a roll-over in the high energy data above $\sim$10 keV. We therefore tried a power-law model with a high-energy exponential cut-off ($cutoffpl$ in XSPEC) to test if it accounted for the high-energy residuals. This model, with an e-folding energy of the cut-off of 20$^{+20}_{-7}$ keV, further improves the fit, with $\Delta \chi^2$=12 for one additional degree of freedom. The cut-off energy, being very close to the high end of the spectral data, is not tightly constrained. 

In order to assess the significance of the potential cutoff,
we performed a series of spectral simulations. Using the
same responses, background files and adopting the same
exposure times as the data used here, we simulated 10,000
sets of \xmm\ (pn, MOS1, MOS2) and \nustar\ (FPMA, FPMB)
spectra with the FAKEIT command in XSPEC. These simulations
were based on the best-fit power-law continuum, and each of
the simulated datasets was rebinned in the same manner and
analyzed over the same bandpass as adopted for  the real
data. We then fit each of the combined datasets with an
absorbed power-law continuum with and without an exponential
cutoff, again in the same manner as the real data, and noted
the improvement in fit provided by the former over the
latter. Of the 10,000 datasets simulated, only 28 showed a
chance improvement equivalent to or greater than that
observed, implying that the detection significance of the
cutoff in the real data is at the 3$\sigma$ level. Next, we performed another set of simulations with the best fit cutoff power-law model as input, in order to assess the expected distribution of $ \Delta \chi^2 $ when compared to a simple power-law model given the available data. We found that it peaks near the observed value and this holds true even if we take into account that the power-law model already provides a good fit with reduced $\chi^2$$<$0.9. All of these simulations favor the detection of a cutoff with 3$\sigma$ significance in our broad-band \xmm+\nustar\ data. Such a feature has been measured for the first time in V404 Cyg during quiescence. 

\begin{figure}
	\centering
	\includegraphics[scale=0.4]{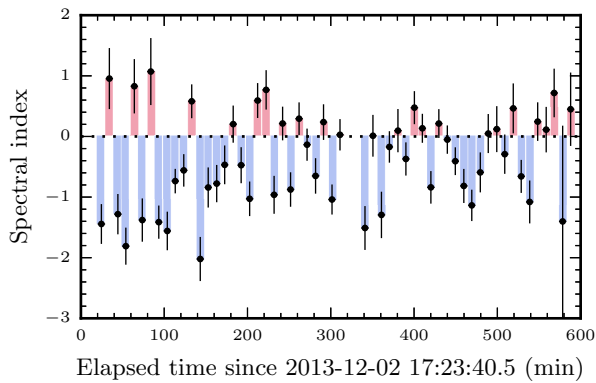}
	\caption{Variation of the spectral index $\alpha$ over the length of the observation obtained by fitting a power-law ($f_{\nu} \propto \nu^{\alpha}$) to the data split in $\sim 10\,$min time interval.\label{vla_spec_variability}}
\end{figure}

\begin{table}
	\centering
	\caption{Summary of V404~Cyg Stokes $I$ measurements. A point source is fitted to each frequency-dependent image integrated over a $512\,$MHz bandwidth. \label{vla-param}}
	\begin{tabular}{ccc}
		\hline \hline
		Frequency & Flux density & Beam size \\ 
		(GHz) & ($\mu$Jy) & (arcsec$^2$) \\ 
		\hline
		$5$ & $314.7 \pm 6.2$ & $1.25 \times 1.21$\\  
		$5.5$ & $307.2 \pm 5.5$  & $1.13 \times 1.09$ \\ 
		$7.2$ &  $282.8 \pm 4.0$ & $0.84 \times 0.83$\\  
		$7.7$ & $281.8 \pm 4.0$ & $0.81 \times 0.79$ \\ 
		\hline 
	\end{tabular} 
\end{table}

We next used a physically motivated thermal bremsstrahlung model to characterize the broad-band spectra. Again, this model provides a statistically acceptable fit to the data. The derived plasma temperature is 6.9$^{+0.7}_{-0.6}$ keV. However, this fit is poorer by $\Delta \chi^2$=33 compared to the cut-off power-law model. Thus the cut-off power-law is the favored spectral model for these data among the above three models.
In addition, we also tried a thermally Comptonized continuum model, $comptt$, that describes Comptonization of soft photons in a hot plasma from a Comptonizing corona and the inner region of an ADAF flow \citetext{\citealp{titarchuk95, hua95}}. The $comptt$ model has a higher number of free parameters, and not all of them are constrained simultaneously. The seed photon temperature $kT_0$ varies between 0 to 0.4 keV with a best fit value of 0.32 keV, which is in excellent agreement with the previous measurement of 0.33 keV by \cite{bernardini14}. Hence we fixed the seed photon temperature to 0.32 keV during the fit. As listed in Table~\ref{spec-param}, the $comptt$ model is characterized by a coronal electron temperature of $\sim$4 keV and an optical depth of $\sim$5. The goodness of fit with this Comptonization model is very similar to that found for the simple cutoff power-law model (see Table~\ref{spec-param}) and shows very similar residuals. Therefore, we have not shown it in Figure~\ref{3model_ratio} for the sake of clarity.

Visual inspection of the X-ray spectra indicates that the data are featureless, showing only a smooth continuum over the broad X-ray band. However, there are predictions for the presence of X-ray emission lines from the hot plasma in an ADAF-like flow  \citep{narayan99}. In order to derive an upper limit on the highly ionized Fe~XXV K$_{\alpha}$ line at 6.7 keV, we added a Gaussian component to our best fit cutoff power-law model. We obtained an upper limit on the equivalent width of $\sim$200 eV (0.1 keV fixed width and 90\% confidence limit)  from the broad-band \xmm\ and \nustar\ spectra. This is somewhat higher than the upper limit derived by \cite{bradley07}. However, it is nearly as large as the prediction of $\sim$230 eV for the sum of the equivalent widths for the total group of Fe emissoin lines \citep{narayan99}.  

\subsection{Radio Spectral Analysis}
In order to characterize the spectral energy distribution (SED) at radio energies, we measured the flux density of V404~Cyg through four distinct frequency bands over the whole observing period. We report the results in Table~\ref{vla-param}.
We use this information to constrain the radio spectral shape by fitting the function $f_{\nu} \propto \nu^{\alpha}$, where $f_v$ is the flux density and $\nu$ is the frequency. Figure~\ref{vla-spec} shows the radio spectrum (black points) fitted with a power-law model (dotted line) and the corresponding 95\% confidence region (gray shaded area).
We obtain a spectral index of $\alpha = - 0.27 \pm 0.03$  (1$\sigma$ error) for the average spectrum across the whole observation.
The same fitting procedure was applied to two frequency bands to derive the evolution of the spectral shape during the observation. Figure~\ref{vla_spec_variability} shows the spectral index variation over $\sim 10\,$min time bins. Interestingly, the spectral index changes rapidly between optically thick (0.0 to 0.5) and thin (-1.0 to -0.5) synchrotron emission regimes. 
In order to make sure that such extreme variability is associated with V404~Cyg, we performed a similar analysis on nearby field sources. We found that the flux density and the derived radio spectral index of the field sources were constant over the course of the observation, confirming that the observed spectral variability displayed in Figure~\ref{vla_spec_variability} is associated with V404~Cyg.

\section{Discussion}
We obtained simultaneous multi-wavelength data for V404~Cyg during its quiescent state, allowing us to extend such studies beyond 10 keV for the first time.  

\subsection{Comparison with Previous Studies}
Over the course of our observations, V404~Cyg showed strong variability at soft X-ray, hard X-ray and radio wavelengths, represented as multiple flares with each flare lasting for a few hours. Similar temporal variability has been observed previously in long term monitoring of this source with $Swift$ XRT \citep{bernardini14} in the 0.3--10 keV energy band. That study showed that V404~Cyg is highly variable on a wide range of time-scales from tens of minutes to years in 34 $Swift$ observations of the source in quiescence. \citet{hynes04} also reported dramatic X-ray flux variability with much higher amplitude (more than a factor of 20) using 2003 $Chandra$ observations in the 0.3--7 keV band. Our sensitive \nustar\ observations clearly demonstrated that the flaring behavior of V404~Cyg continues beyond 10 keV. In addition, new \vla\ radio light curve, taken simultaneously with the \nustar\ hard X-ray data, also showed strong variability on short time scales, similar to that reported by \citet{hynes09}. However, unlike these previous radio light curves, our new radio data show a persistent emission component with the flux level never going to zero. Although the simultaneous radio and X-ray data show strong variability, we do not find any significant correlation (or anti-correlation) between the two data sets. This is again consistent with previous multi-wavelength studies \citep{hynes09}. 

\begin{figure}
	\includegraphics[scale=.45]{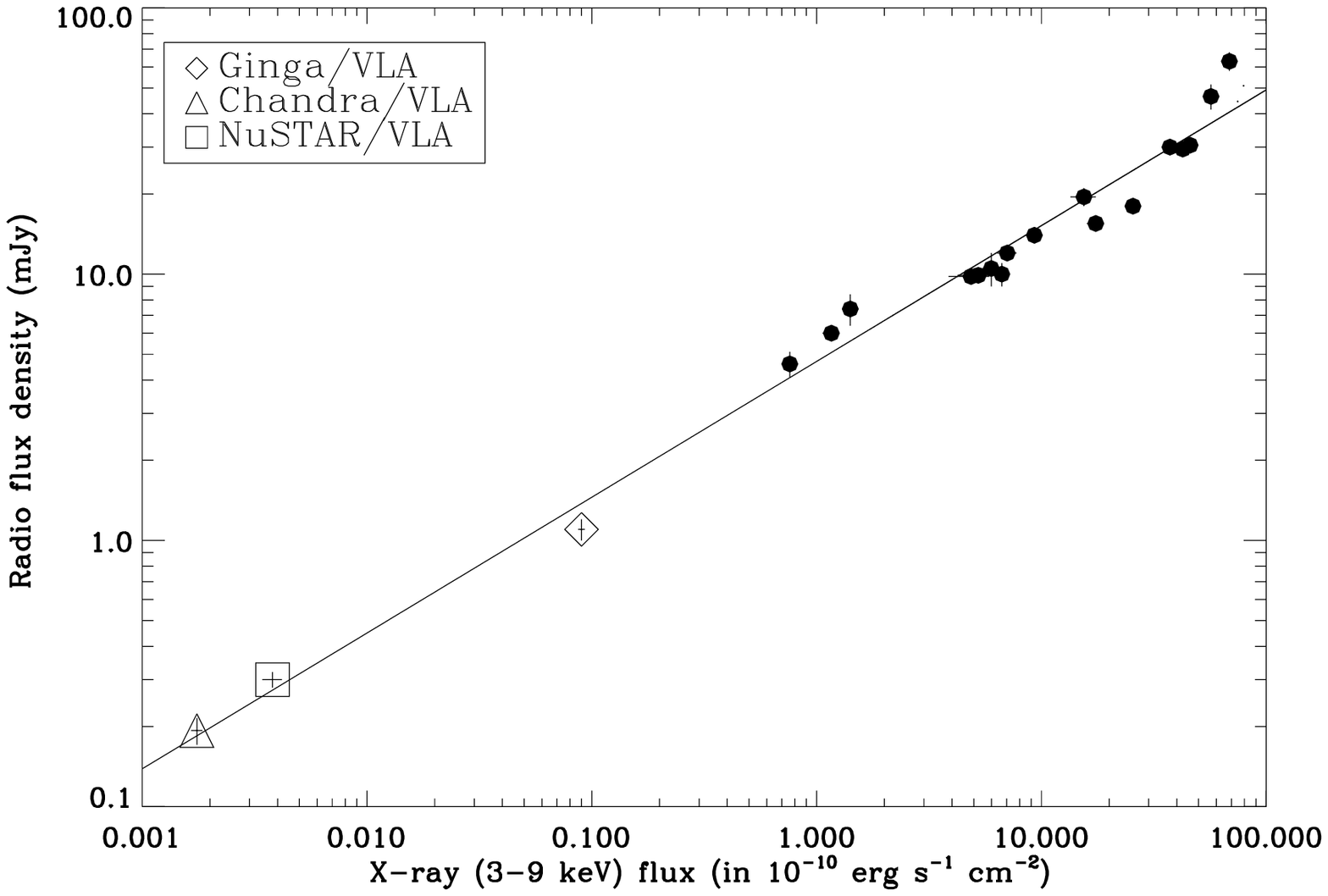}
	\caption{X-ray flux versus Radio flux density variation reproduced from \citet{corbel08} with addition of a new point from our recent \nustar\ and \vla\ observations shown as a square box. Solid circles represent data from 1989 outburst of V404~Cyg.
		\label{radiox-flux}}
\end{figure}

The quiescent X-ray spectra of V404~Cyg have been studied previously with \xmm, \chandra\ and $Suzaku$ \citetext{see \citealp{bradley07}, \citealp{corbel08} and \citealp{reynolds14}} and are well characterized using absorbed power-law models. We measured the power-law index $\Gamma$=2.12$\pm$0.07 using our broad-band \xmm\ and \nustar\ data. This measurement is in excellent agreement with previously reported power-law spectra from \chandra\ ($\Gamma$=2.17$\pm$0.13) and \xmm\ ($\Gamma$=2.09$\pm$0.08). \cite{reynolds14} recently studied quiescent X-ray spectra for a sample of accreting BHs, including V404~Cyg. They reported best fit power-law index of 2.05$\pm$0.06 from a simultaneous fit to \xmm, \chandra\ and \suzaku\ observations, which agrees with our measurement within 90\% errors. The broad-band \xmm\ and \nustar\ spectra, along with excellent sensitivity, show a presence of cutoff at around $\sim$20 keV with more than $\sim$3$\sigma$ significance. A similar cutoff (at around 10 keV) was recently detected by \nustar\ in the quiescent spectra of the low mass X-ray binary Cen X-4 \citep{chakrabarty14} which these authors attribute to hard X-ray emission arising either from a hot, optically thin corona or from hot electrons in an optically thin radiatively inefficient accretion flow (RIAF). The broad-band spectrum of Cen X-4 is well characterized by a thermal bremsstrahlung model with $kT_e$=18 keV. For V404~Cyg, the bremsstrahlung model also provides a statistically acceptable fit, though, the best fit temperature, $kT_e$ = 6.9$^{+0.7}_{-0.6}$ keV, is significantly lower than that observed in Cen X-4. Recently, \citet{bernardini14} analyzed \swift\ XRT spectra of V404~Cyg and reported that fits with a bremsstrahlung model find the temperature between 3.3 to 4.7 keV as a function of count rate, which is slightly lower than our measurement. However, \cite{reynolds14} reported a bremsstrahlung temperature of 6.16$^{+1.15}_{-0.89}$ keV using \suzaku\ XIS data, in agreement with our measurement.

\cite{corbel08} revisited the radio-X-ray flux correlation for V404~Cyg by re-analyzing some of the 1989 outburst data and adding two points from their 2003 \chandra/VLA data and 2000 $Ginga$/VLA observations during quiescence. They found a strong correlation between the two energy bands and conclude that the correlation observed during the decay of the 1989 outburst holds down to the quiescent state. Using our new simultaneous \nustar/\vla\ observations, we were able to add one more point to that correlation, shown in Figure~\ref{radiox-flux}. Earlier data were taken from \cite{corbel08} (see their Figure~4) and we did not attempt any fitting. Our new data point lies right on their best fit curve, nicely following the long-term correlation from previous data. The quiescent flux level of V404~Cyg during our observations is higher compared to their \chandra/VLA observation, but it is significantly lower than the $Ginga$/VLA measurements. This long-term correlation between X-ray and radio flux is indicative of a common or related origin for the two components. Although the radio and X-ray fluxes are correlated on long timescale, the short timescale variations during the observations become more complicated and any correlation may be smoothed out by internal jet variability \citep{gleissner04}.

The quiescent X-ray flux of V404~Cyg shows significant variability on a time scale of a few years, as reported in previous studies. \cite{reynolds14} reported that the 0.3--10 keV unabsorbed X-ray flux varies in the range of (0.8--3.42) $\times$ 10$^{-12}$ erg s$^{-1}$ cm$^{-2}$ using various observations between 2000--2009 from \chandra, \xmm, and \suzaku. We measured the unabsorbed 0.3--10 keV flux to be 1.1 $\times$ 10$^{-12}$ erg s$^{-1}$ cm$^{-2}$ which is well within the above range, however it is slightly lower compared to the average flux of 1.7 $\times$ 10$^{-12}$ erg s$^{-1}$ cm$^{-2}$, reported by \cite{bernardini14} using 75 days of \swift\ monitoring observations in 2012. \cite{narayan97} reported unabsorbed 1--10 keV flux of 8.2 $\times$ 10$^{-13}$ erg s$^{-1}$ cm$^{-2}$ using 1994 $ASCA$ observations. Thus, the X-ray flux of V404 Cyg shows significant variability; however, its spectral shape remains fairly stable with $ \Gamma $=2.1--2.3. 

\subsection{Hard X-ray Emission Mechanisms}
We can use the measured spectral cutoff to make inferences about the nature of
the accretion flow onto the black hole. We follow the general approach
used by \citet{chakrabarty14} in interpreting the spectral cutoff in the
quiescent accreting neutron star Cen X-4.  A key difference here is
that the observed X-ray luminosity in black hole systems provides only a
lower limit on the mass accretion rate, since matter can be accreted
onto the black hole without radiating. In the discussion below, we
assume a black hole mass of 10~$M_\odot$ and a distance of 2.4 kpc,
and we write the Eddington luminosity and accretion rate for a 10
$M_\odot$ black hole as $L_E = 0.1\dot M_E\,c^2 = 1\times 
10^{39}$ erg~s$^{-1}$.  The observed X-ray luminosity is thus
$L_x \simeq 10^{-6}\,L_E$. 

The hard X-ray emission in BH binaries is often ascribed to a
Comptonizing corona of hot electrons.  Our {\tt comptt} fit yielded an
electron temperature of $kT_e = 4.0$~keV and an electron scattering
optical depth of $\tau=4.8$.  This is a fairly high optical depth, and
it is related to the electron density $n_e$ by
\begin{equation}
  \tau = \sigma_T \int n_e(r) dr ,
\end{equation}
where $\sigma_T$ is the Thomson cross-section.  We assume a
quasi-spherical, RIAF inside
a transition radius at $r_t = 10^4 R_S$, where $R_S = 2 GM/c^2 =
30$~km for a 10 $M_\odot$ black hole.  We then find that an 
accretion rate of $\dot M \approx 0.02 \dot M_E$ is required to
produce the measured value of $\tau$. (We used Equations (5) and (8)
of \cite{chakrabarty14} to relate $n_e$ and $\dot M$, taking
$\eta=0.1$ and $\mu=0.6$.)  The resulting electron population has a
high scattering optical depth but is optically thin to free-free
absorption, and will thus itself be a source of bremsstrahlung emission. At a
temperature of $kT=3.9$~keV, this would produce an X-ray luminosity
60 times brighter than what we observe.  If we allow for the
possibility of an outflow by setting
\begin{equation}
        \dot M(r) = \left(\frac{r}{r_t}\right)^p\, \dot M_t
\end{equation}
with $p>0$ \citep{blandford99}, the discrepancy is even
larger. We conclude that thermal Comptonization cannot be responsible
for the spectral cutoff. 

Thermal bremsstrahlung emission from the accretion flow is, however, a
viable possibility.  Our bremsstrahlung fit has an emission measure
\begin{equation}
  \int n_e^2\,dV = 4.7 \times 10^{57} \mbox{\rm\ cm$^{-3}$}
\end{equation}
for a distance of 2.4~kpc. For the $p=0$ case in Equation~(2), an
accretion rate of $\dot M \simeq 10^{-3}\dot M_E$ is sufficient to
yield this emission measure. \citet{menou99} suggested that V404 Cyg during 
its quiescent state is basically in the RIAF regime when mass loss via wind is included. If we allow for an outflow ($p>0$), then we require $\dot M_t\sim 10^{-2} \dot M_E$ at the transition radius. These are plausible accretion rates for the RIAF regime \citep{menou99}.

Another possibility is synchrotron and/or synchrotron self-Compton
emission from the base of a jet outflow \citep{markoff05}. Such
models are competitive with Compton coronal models for the X-ray
emission in black hole binaries and are capable of producing curvature
in the power-law spectrum which could be consistent with our observed
cutoff.  Detailed modeling will be necessary to evaluate the viability
of this model in V404~Cyg.

\subsection{Rapid Radio Spectral Variation}
The hundreds of seconds to ks time-scale flaring behavior of V404~Cyg in quiescence has already been reported at several wavelengths, including at $8.4\,$GHz in the radio \citep{2008MNRAS.388.1751M, hynes09}. Our new sensitive radio observations with the \vla\ confirm this behavior, but they also allow us for the first time to have additional radio spectral information during these periods of weak activity. The spectral index (Figure~\ref{vla_spec_variability}) evolved on a very fast time-scale (as short as $10\,$minutes) with clear switching between regimes of optically thick and optically thin synchrotron emission. 

In order to explain the observed spectral variability, we explored several possibilities, one of them being intervening material along the line of sight. V404~Cyg lies behind the Cygnus super-bubble, composed of interstellar structures shaped by supernova explosions and stellar winds of massive stars, all estimated to be located between $0.5$ and $2.5\,$kpc from the Earth \citep[see e.g.][]{2001A&A...371..675U, 2011Sci...334.1103A}. Fluctuation of the electron density in these clouds can modify the refractive index of the plasma, inducing scintillation of the observed radio emission. However, \citet{2008MNRAS.388.1751M} estimated that the time-scale of refractive scintillation should be between $140$ and $1000\,$hrs, given the source size constraints they derived from VLBI observations. The flux variability seen in our \vla\ observations on few minutes time-scale must therefore be intrinsic to V404~Cyg. 

In the hard state, self-absorbed compact jets are usually observed with a radio spectrum that is either flat or slightly inverted \citep[e.g.][]{2000A&A...359..251C}. However, it has been found that, when the compact jets are building up during the soft to hard state transition, they are initially consistent with optically thin emission \citep[e.g.,][]{2013MNRAS.431L.107C, 2013ApJ...779...95K}. The quiescent state in V404~Cyg may be similar to such a situation with the accretion flow not being able to sustain stable compact jets due either to insufficient particle density and/or inefficient particle acceleration. One could possibly interpret the rapid spectral index changes in Figure~\ref{vla_spec_variability} as jet ignition instabilities, rapidly moving between development/fading ($\alpha < 0$) and activated states (corresponding to flat/inverted spectra, $\alpha \geq 0$). A dedicated campaign with simultaneous observations in infrared and radio should provide an important diagnostic for such behavior. 

Alternatively, the observed fast spectral modulations of the radio spectrum may originate from stochastic instabilities in the accretion flow. The synchrotron turnover frequency, $\nu_b$, of a compact jet \citep[usually at infrared frequencies in the hard state, e.g.][]{ 2002ApJ...573L..35C, 2006MNRAS.371.1334R, hynes09, 2013MNRAS.431L.107C} could move to lower frequencies as the mass accretion rate, $\dot{m}$, decreases \citep{2004A&A...414..895F, 2011A&A...529A...3C}. 
In quiescence, the accretion rate in V404~Cyg may fluctuate on very short time-scales. Rapid modulations of the magnetic field intensity and/or the size of the particle acceleration zone may cause $\nu_b$ to shift over to the range of our observable bandwidth \citep[][]{2011A&A...529A...3C, 2011ApJ...740L..13G}, thus producing the switches between optically thick and thin synchrotron emission, as observed at infrared wavelengths on similar time-scales for other sources \citep{2012MNRAS.422.2202R, 2011ApJ...740L..13G}. The fact that the new observations lie on the standard radio and X-ray flux correlation observed in the hard and quiescence states probably indicates an interpretation related to the compact jets. As observed in \citet{2011ApJ...740L..13G} and \citet{2012MNRAS.422.2202R}, even if the compact jets are strongly variable on short time-scales, those variations smooth out on longer time-scales.
    
The observed radio emission may also be unrelated to compact jets, which might be absent in quiescence. The radio flaring behavior might instead be related to synchrotron bubble ejection events, possibly triggered by magnetic reconnection events above the accretion disk, or be related to variations in the accretion rate.  In the synchrotron bubble scenario, the high-frequency peak should lead the lower low-frequency flux maximum \citep{1966Natur.211.1131V, 1988ApJ...328..600H}. 
Considering averaging effects, this is consistent with the rapid switch between the observed positive and negative  spectral indices (see Figure~\ref{vla_spec_variability}). However, in such a case, one would expect the higher frequency light-curve to have higher maxima than observed at lower frequency. This is in contrast to what we observe (Figure~\ref{vla-lc}), and  indicates that the synchrotron bubble scenario is not the favored interpretation.

\section{Summary}
V404~Cyg shows strong variability on time-scales of hours to a couple of hours, at both X-ray and radio energies. Several small flares were detected at X-ray energies by both \xmm\ and \nustar, during which the count rate changed by a factor of 4--5 in a few hours. On the other hand, the radio flares are comparatively short lived and asymmetric, with fast rises changing flux density by a factor of $\sim$7 in 5 min followed by a slower decay.

Our broad-band X-ray spectral analysis using absorbed power-law models provides consistent results with previous studies using just X-ray data below 10 keV. However, with NuSTAR extending the spectral coverage
above 10 keV, we were able to measure a cut-off at 20$^{+20}_{-7}$ keV, which can be explained using thermal bremsstrahlung emission from the accretion flow or synchrotron emission from the base of a jet outflow. V404~Cyg is  the second quiescent LMXB showing such a feature, in addition to the neutron star system Cen X-4 \citep{chakrabarty14}.

Our \nustar\ and \vla\ observations are in very good agreement with the long-term correlation between X-ray and radio flux measured over many orders of magnitude \citep{corbel08}. This  indicates that the origins of these two components are not completely independent on long timescales. 

Our new sensitive radio observations discovered extreme variability in the spectral index on time-scales as short as 10 minutes, with clear switching between optically thick and thin synchrotron emission. Such variability could possibly be attributed to instabilities in the compact jet or stochastic instabilities in the accretion rate where the synchrotron turnover frequency could shift to lower energies during quiescence. 

\acknowledgments
We thank Michael A. Nowak for useful discussion on synchrotron model.
This work was supported under NASA contract No. NNG08FD60C, and made use of data from the \nustar\ mission, a project led by the California Institute of Technology, managed by the Jet Propulsion Laboratory, and funded by NASA. We thank the \nustar\ Operation, Software and Calibration teams for support with the execution and analysis of these observations. This research has made use of the \nustar\ Data Analysis Software (NuSTARDAS) jointly developed by the ASI Science Data Center (ASDC, Italy) and the California Institute of Technology (USA). SC would like to thank James Miller-Jones for excellent help during the preparation of the \vla\ observations. SC and AL  acknowledge the  financial support from the UnivEarthS Labex
programme of Sorbonne Paris Cit\'e (ANR-10-LABX-0023 and ANR-11-IDEX-0005-02), and from the CHAOS project ANR-12-BS05-0009 supported by the French Research National Agency. JAT acknowledges partial support from NASA under {\em XMM-Newton} Guest Observer grant NNX14AF08G.



{\it Facilities:} \facility{\nustar}, \facility{\xmm}, \facility{\vla}.


\bibliography{xrbs}

\end{document}